\pdfoutput=1
\documentclass[12pt]{iopart}

\usepackage{epsfig}
\usepackage[utf8]{inputenc}

\usepackage{mathptmx}
\usepackage[scaled=.90]{helvet}
\usepackage{courier}

\newcommand{\text}[1]{\textrm{#1}}
\newcommand{\vg}{\ensuremath{V_\text{g}}}
\newcommand{\vsd}{\ensuremath{V_\text{sd}}}
\newcommand{\idc}{\ensuremath{I_\text{dc}}}
\newcommand{\un}[1]{\ensuremath{\,{\rm{#1}}}}

\newcommand{\nume}{\ensuremath{N_{\text{e}^-}}}

\begin{document}

\title{Magnetic damping of a carbon nanotube NEMS resonator}

\author{D R Schmid, P L Stiller, Ch Strunk and A K Hüttel}
\address{
Institute for Experimental and Applied Physics, University of
Regensburg, \\
93040 Regensburg, Germany
}
\ead{andreas.huettel@physik.uni-regensburg.de}

\begin{abstract}
A suspended, doubly clamped single wall carbon nanotube is characterized at
cryogenic temperatures. We observe specific switching effects in dc-current
spectroscopy of the embedded quantum dot. These have been identified previously
as nano-electromechanical self-excitation of the system, where positive
feedback from single electron tunneling drives mechanical motion. A magnetic
field suppresses this effect, by providing an additional damping mechanism.
This is modeled by eddy current damping, and confirmed by measuring the
resonance quality factor of the rf-driven nano-electromechanical resonator in an
increasing magnetic field.
\end{abstract}


\maketitle

Nano-electromechanical resonator systems provide an intriguing field of
research, where both technical applications and fundamental insights into the
limits of mechanical motion are possible. Among these systems, carbon nanotubes
provide the ultimate electromechanical beam resonator \cite{nature-sazonova:284,
nl-witkamp:2904, prl-reulet:2829}, because of their stiffness, low mass,
and high aspect ration. At the same time, they are an outstanding material for
transport spectroscopy of quantum dots at cryogenic temperatures
\cite{science-bockrath:1922,nature-tans:474}. Chemical vapour
deposition (CVD) has been shown to produce on chip defect-free single wall
carbon nanotubes \cite{nature-kong:878}. By performing this growth process as
last chip fabrication step, suspended defect- and contamination-free
macromolecules can be integrated into electrode structures and characterized. On
the electronic side, this has led to many valuable insights into, e.g., the
physics of spatially confined few-carrier systems \cite{nmat-cao:745,
nature-kuemmeth:448, nnano-steele:363}. In terms of nano-electromechanical
systems, these ultra-clean nanotubes have shown exceedingly high mechanical
quality factors at cryogenic temperatures \cite{highq}. This has allowed for
the observation of direct interaction between single electron tunneling and
mechanical motion \cite{strongcoupling, pssb-huettel, science-lassagne:1107}.

\begin{figure*}[t]
\centering
\epsfig{file=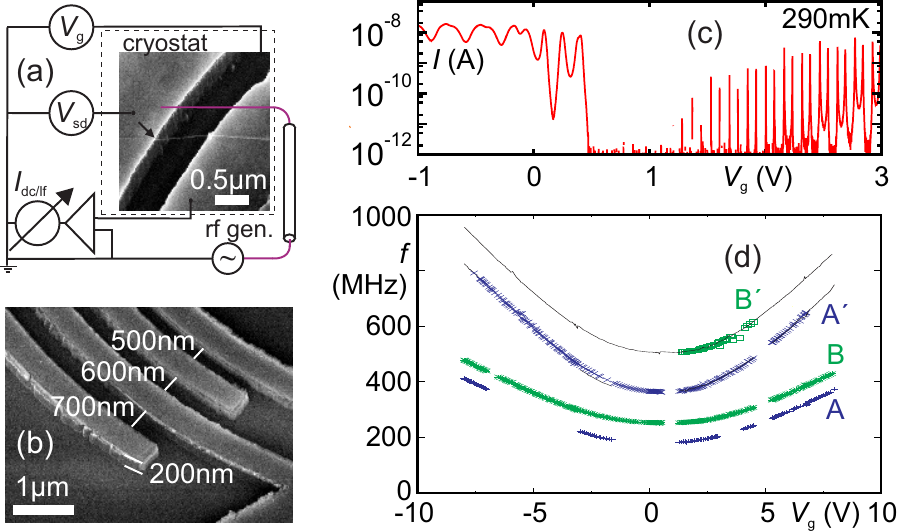,width=0.8\textwidth}
\caption{\label{fig-sample} Measurement setup, chip geometry and basic sample
characterization. (a) Scanning electron microscope (SEM) image of a typical
suspended carbon nanotube, combined with a simplified sketch of the measurement
electronics. dc gate and bias voltages are applied and the low-frequency current
signal measured via a preamplifier; in addition a radio frequency signal can be
applied via an antenna hanging several centimeters above the chip. (b) SEM image
of the electrode geometry, displaying concentric, ring segment shaped contact
electrodes
around a nanotube growth catalyst island. All transport measurements presented
here are measured across the same $700\un{nm}$ wide trench. (c) Measurement of
the dc current $\idc(\vg)$ through the nanotube device as function  of back gate
voltage \vg\ for an applied bias $\vsd=0.2\un{mV}$. (d) Observed rf-driven
mechanical resonance frequencies $f(\vg)$ as function of back gate voltage \vg,
see text for details.
}
\end{figure*}
In this article, we report on low temperature transport spectroscopy
measurements on a suspended, doubly clamped carbon nanotube, as displayed in
figure~\ref{fig-sample}(a). The carbon nanotube acts as an ultra-clean quantum
dot as well as a nano-electromechanical transversal resonator.
Figure~\ref{fig-sample}(b) shows a typical chip electrode structure including
dimensions. On a highly p+ doped Si substrate with $\sim 300\un{nm}$ thermally
grown $\text{SiO}_2$ on top, contact patterns are defined via electron beam
lithography and evaporation of $40\un{nm}$ rhenium. This metal layer directly
serves as etch mask for subsequent anisotropic dry etching of the oxide,
generating deep trenches between the electrodes. As last fabrication step, CVD
growth catalyst is locally deposited at the center of each contact electrode
structure and the nanotube growth is performed \cite{nature-kong:878}. 

Electronic transport measurements were conducted in a ${}^3\rm{He}$ evaporation
cooling system at $T_{{}^3\rm{He}}=290\un{mK}$, and in a dilution refrigerator
at $T_\text{mc,base} = 25\un{mK}$. The electronic measurement setup, as sketched
in figure~\ref{fig-sample}(a), closely follows Refs.~\cite{highq,
strongcoupling}. A gate voltage \vg\ is applied to the substrate as back
gate, a bias voltage \vsd\ across the device. The resulting dc current
through the device is measured via a preamplifier, as required for Coulomb
blockade transport spectroscopy \cite{kouwenhoven}. An antenna suspended close
to the chip provides means to apply a radio-frequency signal contact-free.

At first, we characterize the basic electronic and electromechanical properties
of the device. As can be seen from the dc current measurement in
figure~\ref{fig-sample}(c), our device exhibits the typical electronical
behavior of a very clean and regular small band gap nanotube. The measurement
displays the dc current $I_\text{dc}$ as function of the applied gate voltage
\vg, for a low constant dc bias voltage $\vsd=0.2\un{mV}$. For $\vg <
0.5\un{V}$, highly transparent contacts in hole conduction lead to a rapid
transition from Coulomb blockade to the Fabry-Perot interference regime
\cite{nature-liang:665}. Around $\vg \simeq 0.75\un{V}$, current is suppressed
as the electrochemical potential is located within the semiconducting band gap.
For $\vg > 1\un{V}$, electron conduction becomes visible through sharp,
well-defined Coulomb blockade oscillations with the characteristic four-fold
pattern of the carbon nanotube level structure \cite{prl-oreg:365}. Regular
Kondo conductance enhancement \cite{nature-goldhaber:156} emerges for
$\nume>15$, again confirming the presence of a defect-free single wall carbon
nanotube.

When a radiofrequency signal is applied at mechanical resonance, the nanotube
vibrates,
leading to a change in detected, time-averaged dc current \cite{highq}. This
signal can be identified via its characteristic dependence on the back gate
voltage \vg: electrostatical forces on the influenced charge on the nanotube
lead to mechanical tension, and thereby an increase in resonance frequency of
the transversal vibration mode. Figure~\ref{fig-sample}(d) shows a map of such
resonance positions, displaying the resonance frequency as function of back gate
voltage \vg. It thus characterizes the basic electromechanical properties of
our device. Among several other weaker features, four clear structures, plotted
in figure~\ref{fig-sample}(d) and labelled A, B, A', and B', can be seen in the
observed frequency range, with the overall gate voltage dependence typical for
the mechanical response of a tensioned carbon nanotube resonator
\cite{nature-sazonova:284, nl-witkamp:2904, highq}. 

Traces A' and B' coincide over a wide range with double the frequency of traces
A and B (plotted in
figure~\ref{fig-sample}(d) as thin black lines). It appears unlikely that these
represent higher mechanical modes, since in the low tension limit an exact
frequency doubling is not expected \cite{book-cleland}. Instead, A' and B' can
represent different driving mechanisms for the modes of A and B. In literature,
e.g., parametric resonance has been demonstrated in measurements on nanotube
resonators \cite{nl-eichler:x,nl-laird}. The observation of the two modes A and
B is consistent with mechanical motion of two adjacent suspended nanotube
segments of different length, as visible in the chip geometry of
figure~\ref{fig-sample}(b). Assuming the minimum resonance frequency close to
charge neutrality to be the case of vanishing mechanical tension, the ratio of
the minimum frequencies of A and B, $f_\text{min,A} / f_\text{min,B} =
253\un{MHz} / 182\un{MHz} \simeq 1.39$ agrees very well with the expectation
from the different trench widths $(\ell_B / \ell_A)^2 = (700\un{nm} /
600\un{nm})^2 \simeq 1.36$. The detailed mechanism leading to the signal
contribution of the second nanotube segment next to the contacted $700\un{nm}$
gap is still under investigation.

\begin{figure*}[t]
\centering
\epsfig{file=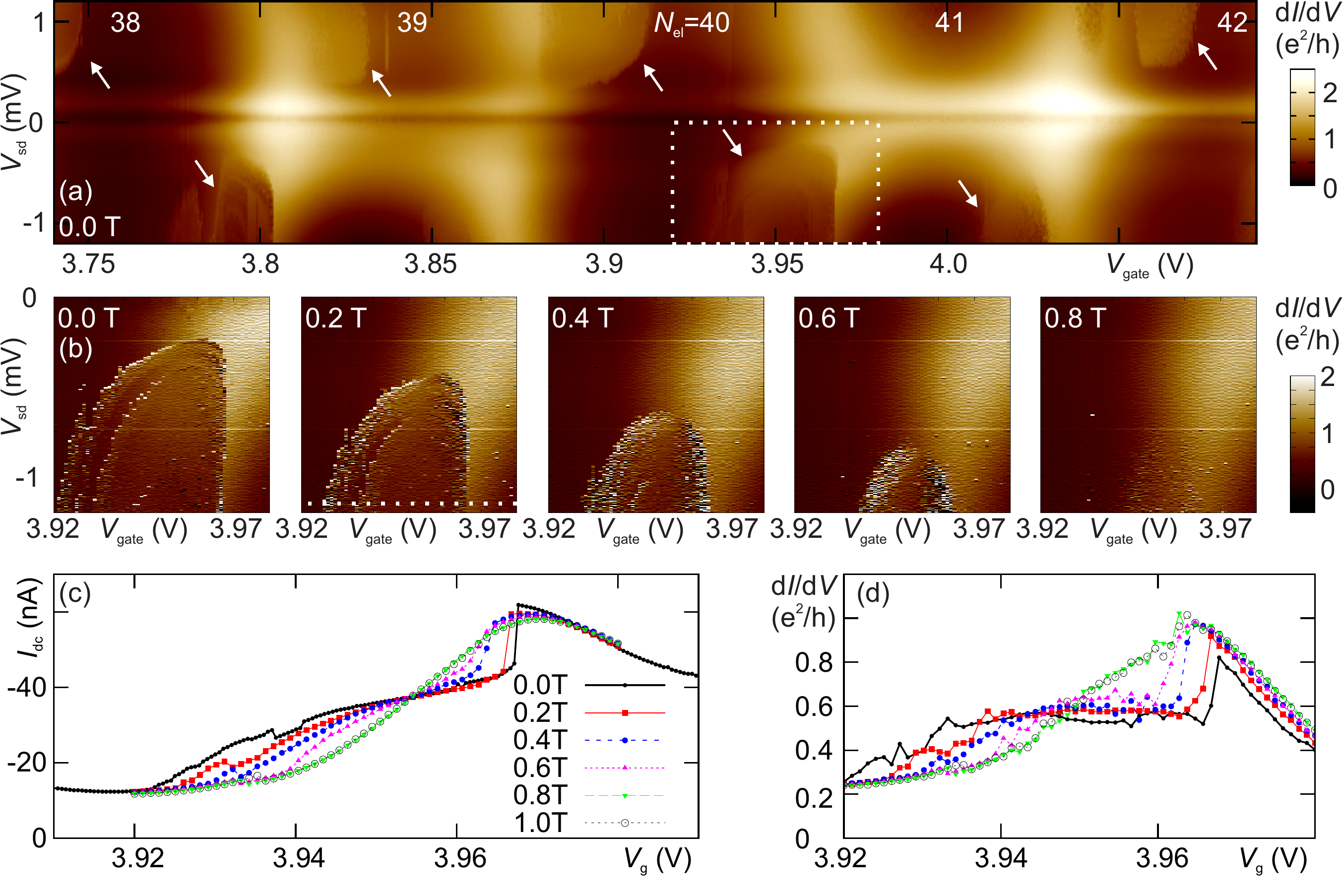,width=\textwidth}
\caption{\label{fig-switching}
Feedback effects in a non-driven resonator (all measurements at
$T_{\text{mc,base}}$). (a) Differential conductance $\text{d}I / \text{d} \vsd
(\vg, \vsd)$ (no rf signal applied; lock-in measurement with an excitation of
$V_\text{sd,ac} = 5\,\mu\text{V}$ RMS at $137\un{Hz}$) of the carbon nanotube
quantum dot at zero magnetic field, displaying four-fold shell filling combined
with Kondo effect and traces of superconductivity in the metallic leads (see
text). At finite bias, strong switching effects attributable to mechanical
self-excitation become visible, indicated by white arrows \cite{strongcoupling,
Usmani2007Strong}. (b) Detail of (a) for increasing magnetic field, this time
plotting the numerical derivative of the simultaneously measured dc current
$I_{\text{dc}}(\vg, \vsd)$. Already at $B=0.8\un{T}$, the self-driving effects
are completely suppressed. (c) dc current $I_{\text{dc}}(\vg)$ along the trace
of constant $\vsd=-1.15\un{mV}$, as marked in (b) with a dotted line, and (d)
corresponding differential conductance $\text{d}I / \text{d}\vsd(\vg)$.
}
\end{figure*}
In the following we return to measurements {\em without} any applied rf driving
signal. Figure~\ref{fig-switching}(a) displays a lock-in measurement of the
differential conductance of the suspended carbon nanotube quantum dot, as
function of gate voltage \vg\ and bias voltage \vsd. A positive gate voltage is
used to tune the quantum dot into the regime with an electron number of $38\le
\nume \le 42$, where it is strongly tunnel-coupled to the contact electrodes.
Several features on the plot can immediately be identified and are well
understood. The narrow, approximately gate voltage independent conductance
minimum around $\vsd=0$ in figure~\ref{fig-switching}(a) is caused by
superconductivity of the metallic rhenium leads; two Kondo ridges of enhanced
low-bias conductance become clearly visible around $\vg=3.84\un{V}$ ($\nume=39$)
and $\vg = 4.0\un{V}$ ($\nume=41$). 

In addition, the differential conductance signal from
figure~\ref{fig-switching}(a) exhibits sharply delineated regions of modified
signal level, often accompanied by switching behavior, see white arrows in the
figure. This has already been observed previously in clean suspended carbon
nanotube quantum dots \cite{strongcoupling}. As predicted in
Refs.~\cite{Usmani2007Strong, phd-usmani} and confirmed in
Ref.~\cite{strongcoupling}, in these parameter regions single electron tunneling
from dc current alone suffices to coherently drive the mechanical motion via a
positive feedback mechanism. In turn, this becomes visible in the recorded
current or conductance signal as well.

The panels of figure~\ref{fig-switching}(b) display a detail enlargement of the
parameter region of figure~\ref{fig-switching}(a), this time plotting as
differential conductance the numerical derivative of the dc current recorded
simultaneously with the lock-in signal. Although this value is affected by a
larger noise level, it reproduces more faithfully one-time switching events
while sweeping the bias voltage, which delimit the feedback regions. A clear
substructure emerges inside the feedback region, which so far has not found any
equivalent in theoretical considerations. In addition, when increasing an
externally applied magnetic field perpendicular to the chip surface, the
parameter regions of positive feedback shrink. As can be seen from the panels of
figure~\ref{fig-switching}(b), applying a magnetic field of $B=0.8\un{T}$
already
completely suppresses the self-driving effect within the observed region. 

This is further illustrated by the line traces of figure~\ref{fig-switching}(c)
and (d), displaying the dc current (c) and the differential conductance (d) as
function of gate voltage \vg\ at constant $\vsd=-1.15\un{mV}$ across the
parameter regions of figure~\ref{fig-switching}(b). While no significant changes
take place outside the positive feedback region, the discontinuous behavior at
zero field becomes smooth, and only slight fluctuations remain at $B=1\un{T}$.
In particular the current agrees very well with the prediction of
Refs.~\cite{Usmani2007Strong,phd-usmani} in the cases of present and suppressed
feedback.

While it has been shown in Ref.~\cite{strongcoupling} that large electronic
tunnel rates are an important prerequisite for self-excitation, here the
conductance remains unchanged outside the feedback-dominated regions. The
magnetic field does not significantly influence the electronic tunnel rates,
excluding such a mechanism for the suppression of the self-excitation. A second
prerequisite is a high mechanical quality factor \cite{Usmani2007Strong,
phd-usmani}, since the feedback mechanism has to compensate and overcome damping
of the mechanical oscillation. Consequently, the suppression of self-driving
indicates a magnetic field induced additional damping mechanism.

To verify this conclusion from the pure dc measurements, we measure the
frequency dependence of the {\em radio frequency-driven} resonator response. An
additional
damping mechanism in a magnetic field should here become visible as a resonance
peak broadening, i.e. a decrease in effective quality factor $Q$. A constant
positive gate voltage $\vg=3.91\un{V}$ is used to tune the quantum dot into the
Coulomb blockade region with electron number $\nume=40$. Because of the
transparent tunnel barriers to the leads, significant cotunneling conductance on
the order of $G_\text{cot}\simeq 0.4\,e^2/h$ can still be observed in this
parameter region, enabling the detection of the mechanical resonance in dc
current. Extending the mechanical resonance detection setup of
Refs.~\cite{highq,strongcoupling} to increase sensitivity, the applied radio
frequency signal is amplitude-modulated at a low frequency $f_\text{am} =
137\un{Hz}$, such that the period $1/f_\text{am}\simeq 7\un{ms}$ is larger than
the oscillation decay timescale $\sim Q/f$ expected from literature
\cite{highq}. The corresponding low-frequency modulation of the current signal
is recorded by a lock-in amplifier. In addition, we drive at double frequency
(A' in figure~\ref{fig-sample}(d)) as this results in a stronger resonance
signal.

\begin{figure*}[t]
\centering
\epsfig{file=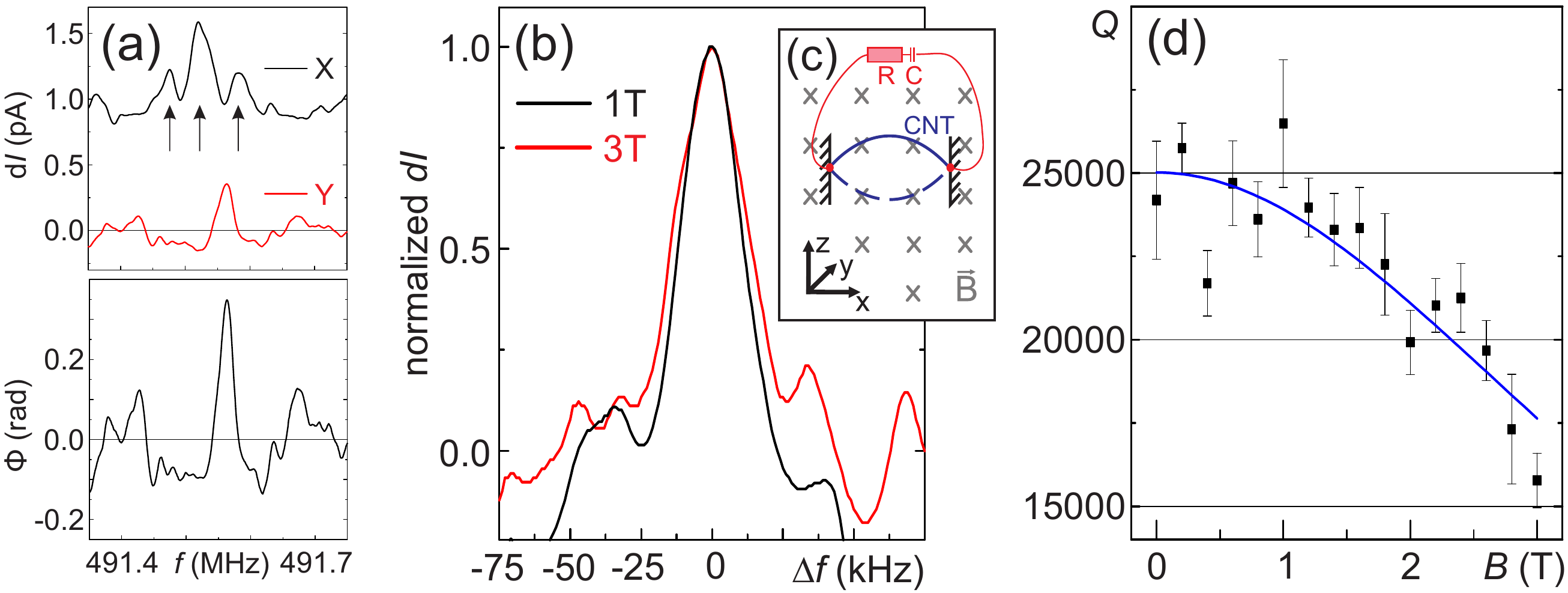,width=\textwidth}
\caption{\label{fig-resonance}
Broadening of the driven mechanical resonance in a magnetic field (all
measurements at $T_{\text{mc,base}}$). (a) Observed mechanical resonance for
$\vg=3.91\un{V}$ and $\vsd=0.3\text{mV}$ (nominal maximal driving signal after
attenuators $-42\un{dBm}$). The mechanical driving signal is amplitude-modulated
at low frequency $f_\text{am}=137$Hz; the plot displays the lock-in response
(upper panel: $x$, $y$, lower panel: $\phi$) at that frequency of the averaged
current through the carbon nanotube quantum dot \cite{highq}. (b) Selected
mechanical resonance curves ($y$ lock-in signal) for $B=1\un{T}$ and $B=3\un{T}$
(nominal maximal driving signal $-41\un{dBm}$). At the higher magnetic field, a
slightly broader resonance can be observed. (c) Circuit model of
electromechanical damping by Ohmic dissipation, see text. (d) Quality factor $Q$
of the resonance extracted from multiple traces as in (b), as function of
magnetic field. The solid line provides a model fit, assuming both a magnetic
field induced and a magnetic field independent damping contribution (see text).
}
\end{figure*}
Figure~\ref{fig-resonance}(a) displays a typical resulting in-phase ($x$),
out-of-phase ($y$), and phase angle ($\phi$) amplitude modulation response
signal as function of the driving frequency $f$. In the in-phase ($x$) response,
a multi-peak structure emerges. Indications of this multi-peak shape (see arrows
in figure~\ref{fig-resonance}(a)) remain visible even at lowermost driving power
and suggest a more complex coupling of the radiofrequency driving signal into
the electromechanical system than only actuation via electrostatical
force \cite{highq,nl-laird,nl-eichler:x}. As can be seen from both the $y$
response and the phase angle, in spite of the low amplitude modulation
frequency, a distinct phase shift of the response on resonance is still visible.
The shift of approximately $\Delta \phi=0.2\un{rad}$ corresponds to a delay time
of $\Delta t=0.35\un{ms}$, on the order of $10^5$ mechanical oscillation cycles.
Given that previously observed nanotube resonators \cite{highq, nl-laird} have
exhibited quality factors on that order of magnitude, this is consistent with
mechanical storage of vibration energy and later release within one amplitude
modulation cycle, leading to a delayed driving response.

To avoid fitting of the multi-peak structure in the in-phase ($x$) signal, we
focus in the following on the out-of-phase ($y$) signal induced by the phase
shift. Figure~\ref{fig-resonance}(b) shows selected frequency response traces
of the mechanical resonance, recorded at an external magnetic field of
$B=1\un{T}$ and $B=3\un{T}$, respectively, and pointing towards a slight
broadening of the peak structure at higher magnetic field. Evaluation of many
similar curves,
including repeated measurements at the same magnetic field value and over a
large driving power range, leads to the plot of figure~\ref{fig-resonance}(d).
Here, the width of the resonance peaks is plotted in terms of an experimentally
observed quality factor $Q$ as function of the magnetic field $B$. Indeed, the
measured peak width increases (and $Q$ decreases) significantly above
$B=1\un{T}$.

A straightforward circuit model sketched in figure~\ref{fig-resonance}(c) can be
used to describe the magnetic field dependence. A vibration component
perpendicular to the magnetic field leads to an induced ac voltage across the
resonator. We assume the carbon nanotube resonator to be partially electrically
shortened in the rf signal frequency range via an Ohmic resistance $R$ and a
large parasitic capacitance. For simplicity, we do not take into account the
deflection shape but assume a uniform beam deflection along the entire nanotube
of length $L$ and mass $m$ to estimate the magnetic flux modulation. As a
result, eddy currents lead to a damping of the mechanical motion corresponding
to
\begin{equation}
 Q_\text{m}(B)= \frac{q}{B^2} \qquad q= 2\pi f \frac{R m}{L^2 2\sqrt{2}},  
\end{equation}
as both induced voltage and resulting eddy current are proportional to $B$.
Assuming an additional magnetic field independent resonator damping, which
determines the zero external field quality factor $Q_0$, we obtain the
expression
\begin{equation}
 Q(B)=\frac{Q_0 Q_\text{m}(B)}{Q_0 + Q_\text{m}(B)}
\end{equation}
The solid line in figure~\ref{fig-resonance}(d) provides a best fit of this
function to the data, using $Q_0$ and $q$ as free parameters and resulting in
the values $q=5.381\times 10^5 \un{T}^2$ and $Q_0=25020$. As visible in
figure~\ref{fig-resonance} this model describes our measurement results well.
This thereby confirms the presence of a magnetic-field induced damping
mechanism. Using the resonance frequency $f$ and estimating $L\simeq
700\un{nm}$ and $m\simeq 1.3\cdot 10^{-21}\un{kg}$, we obtain a value for the
Ohmic resistance of $R\simeq 200\,\text{k}\Omega$ in the replacement circuit
of figure~\ref{fig-resonance}(c). 

As a last remark, using the fit function of figure~\ref{fig-resonance}(d) one
obtains $Q(0.8\un{T}) / Q(0\un{T})\simeq 0.97$, i.e. only a very small decrease
of the effective quality factor within the magnetic field range covered in
figure~\ref{fig-switching}. A likely reason for this is that the resonance peak
widths evaluated in figure~\ref{fig-resonance}(d) do not solely correspond to
the
device quality factor entering the self-excitation, but are broadened by
additional mechanisms, leading to an underestimation of the low-field quality
factor $Q_0$.

Summarizing, we characterize a quantum dot in a suspended ultra-clean single
wall carbon nanotube, which also acts as nano-electromecanical resonator.
We observe how feedback and self-driving effects, where only dc current is
sufficient to drive resonator motion, are suppressed in a finite magnetic
field. The conclusion that the magnetic field induces an additional damping
mechanism is confirmed by tracing the driven resonator response as a function
of magnetic field. We model the decrease of the mechanical quality factor
successfully using eddy current damping.

\ack

The authors would like to thank the Deutsche Forschungsgemeinschaft (Emmy
Noether grant Hu 1808/1, SFB 631 TP A11, GRK 1570) and the Stu\-dien\-stif\-tung
des deutschen Volkes for financial support.

\section*{References}

\bibliographystyle{iopart-num}
\bibliography{paper}

\end{document}